\documentclass[11pt,a4paper]{article}
\usepackage{jheppub_kim}
\usepackage{rotating} 
\usepackage{graphicx,epsfig}
\usepackage{amsmath}
\usepackage {amssymb}
\usepackage{subfigure}

\usepackage{relsize}

\providecommand{\U}[1]{\protect\rule{.1in}{.1in}}

\newcommand{\newc}{\newcommand}

\newc{\be}{\begin{equation}}
\newc{\ee}{\end{equation}}

\newc{\ba}{\begin{eqnarray}}
\newc{\ea}{\end{eqnarray}}
\newc{\bea}{\begin{eqnarray*}}
\newc{\eea}{\end{eqnarray*}}
\newc{\D}{\partial}
\newc{\ie}{{\it i.e.} }
\newc{\eg}{{\it e.g.} }
\newc{\etc}{{\it etc.} }
\newc{\etal}{{\it et al.}}
\newc{\lcdm}{$\Lambda$CDM }

\newc{\ra}{\Rightarrow}

\title{Barrow holographic dark energy with varying exponent}

 \author[a,b,c]{Spyros Basilakos}
  \author[d]{Andreas Lymperis}
 \author[a,e]{Maria Petronikolou}
 \author[a,f,g]{Emmanuel N. Saridakis}

\affiliation[a]{National Observatory of Athens, Lofos Nymfon, 11852 Athens, 
Greece}
\affiliation[b]{Academy of Athens, Research Center for Astronomy and Applied 
Mathematics, Soranou Efesiou 4, 11527, Athens, Greece}

\affiliation[c]{School of Sciences, European University Cyprus, Diogenes 
Street, Engomi, 1516 Nicosia, Cyprus}

\affiliation[d]{Department of Physics, University of Patras, 26500 Patras, 
Greece}

\affiliation[e]{Department of Physics, National Technical University of Athens, 
Zografou Campus GR 157 73, Athens, Greece}

\affiliation[f]{Department of Astronomy, School of Physical Sciences, 
University of Science and Technology of China, Hefei 230026, P.R. China}

 \affiliation[g]{Departamento de Matem\'{a}ticas, Universidad Cat\'{o}lica del 
Norte, 
Avda.
Angamos 0610, Casilla 1280 Antofagasta, Chile}

   \emailAdd{svasil@academyofathens.gr}
\emailAdd{petronikoloumaria@mail.ntua.gr}
\emailAdd{alymperis@upatras.gr}
\emailAdd{msaridak@noa.gr}

\abstract
{We construct Barrow holographic dark energy with varying exponent. Such an 
energy-scale-dependent behavior  is typical in  quantum field theory and 
quantum 
gravity under renormalization group considerations, however in the present 
scenario it has an additional justification, since in realistic cases one 
expects that Barrow entropy quantum-gravitational effects to be stronger at 
early times and to smooth out and disappear at late times. We impose specific, 
redshift-dependent  ans\"{a}tze  for the Barrow running exponent, such as the 
linear, CPL-like, exponential, and trigonometric ones, and we 
investigate their cosmological behavior. We show that we can recover the 
standard thermal history of the universe, with the sequence of 
matter and dark energy epochs, in which the  transition from deceleration to 
acceleration happens at  $z\approx 0.65$, in agreement with observations. In 
the most realistic case of hyperbolic tangent ansatz, in which we can  easily 
bound Barrow exponent inside its theoretically determined bounds 0 and 1 for 
all redshifts,
we see that the dark-energy equation-of-state parameter can be quintessence 
like, or experience   the phantom-divide crossing, while in the future 
it can either tend to the cosmological constant value or   start increasing 
again. All these features reveal that  Barrow holographic dark energy with 
varying exponent is not only theoretically more justified than the standard, 
constant-exponent case, but it leads to richer cosmological behavior too. }

\begin{document}
\maketitle

\section{Introduction}

The progression from the matter era to the accelerated expansion phase in the 
late universe is now a well-established phenomenon. While the cosmological 
constant is the simplest explanation, challenges related to its 
quantum-field-theoretical calculation and the possibility of a dynamical nature 
have prompted two primary approaches to construct  extended scenarios. The 
first retains general relativity as the underlying gravitational theory and 
introduces the concept of dark energy 
\cite{Copeland:2006wr, Cai:2009zp, Bamba:2012cp} as the acceleration source. 
The second involves constructing extended    gravity theories which provide the 
required richer structure \cite{Nojiri:2010wj,Capozziello:2011et, 
Cai:2015emx,CANTATA:2021ktz}.

An alternative explanation for the origin of dark energy emerges through the 
cosmological application of the holographic principle \cite{Fischler:1998st, 
Bak:1999hd, Horava:2000tb}. This framework, rooted in the thermodynamics of 
black holes, connects the ultraviolet cutoff of a quantum field theory   with 
the largest distance of the theory, a prerequisite for its applicability at 
large distances \cite{Cohen:1998zx}. In a region where entropy is proportional 
to volume, the total energy should not exceed the mass of a black hole with the 
same radius, and this saturation leads to the extraction of a holographically 
originated vacuum energy, and thus to a form of holographic dark energy 
with a dynamic nature \cite{Li:2004rb, Wang:2016och}.
The cosmological implications of holographic dark energy are intriguing 
\cite{Li:2004rb, Wang:2016och, Horvat:2004vn,Huang:2004ai, Pavon:2005yx, 
Wang:2005jx,Nojiri:2005pu,Kim:2005at,Wang:2005ph,Setare:2006wh,Setare:2008pc} 
and align with observations 
\cite{Zhang:2005hs,Li:2009bn,Feng:2007wn,Zhang:2009un,Lu:2009iv, 
Micheletti:2009jy}. 
Additionally, one has the advantage that holographic scenarios are free from 
potential pathologies that may appear in   modified gravity   
\cite{CANTATA:2021ktz}. Hence,   holographic dark energy has  spurred 
extensive research, leading to various extensions
\cite{Gong:2004fq,Saridakis:2007cy,Setare:2007we,Cai:2007us,Setare:2008bb,
Saridakis:2007wx,Jamil:2009sq,Gong:2009dc,Suwa:2009gm,BouhmadiLopez:2011xi,
Malekjani:2012bw,Khurshudyan:2014axa, 
Landim:2015hqa,Pasqua:2015bfz,Jawad:2016tne,Pourhassan:2017cba, 
Saridakis:2017rdo,Nojiri:2017opc,Oliveros:2019rnq, Nojiri:2019kkp,
Kritpetch:2020vea,Dabrowski:2020atl,daSilva:2020bdc,Bhattacharjee:2020ixg, 
Lin:2021bxv,Colgain:2021beg, 
Hossienkhani:2021emv,Nojiri:2021iko,Shekh:2021ule,Maity:2022gdy,
Lymperis:2023prf, 
Shaikh:2022ynt}.

One large class of holographic dark energy extension is obtained by changing 
the underlying entropy relation. In particular, the base models 
use the standard Bekenstein-Hawking entropy. However, since there are many 
extended entropies, that arise through various considerations, such as Tsallis 
non-additive entropy \cite{Tsallis:1987eu}, Barrow quantum-gravity corrected 
entropy \cite{Barrow:2020tzx}, Kaniadakis relativistic entropy  
\cite{Kaniadakis:2002zz,Kaniadakis:2005zk}, power-law corrected entropy  
\cite{Das:2007mj,Radicella:2010ss} etc, one can respectively obtain Tsallis 
holographic dark energy \cite{Saridakis:2018unr,Sadri:2019qxt}, Barrow 
holographic dark energy 
\cite{Saridakis:2020zol,Saridakis:2020cqq,Adhikary:2021xym}, Kaniadakis 
holographic dark energy \cite{Drepanou:2021jiv,Hernandez-Almada:2021aiw}, 
power-law holographic dark energy \cite{Telali:2021jju}, etc.

All the above extended-entropy scenarios incorporate a constant parameter that 
quantifies the deviation form standard  entropy. However, in principle one 
expects a dynamical, energy-scale-dependent behavior for this parameter, 
aligning with the typical case observed in quantum field theory and quantum 
gravity under renormalization group applications. In particular, all parameters 
and coupling constants demonstrate a running nature in tandem with the energy 
scale, and in a cosmological framework this would effectively generate a 
time-dependence. A first investigation of Tsallis holographic dark energy models 
with 
extended entropy with varying exponent was performed in \cite{Nojiri:2019skr} 
and it was shown that the running behavior can lead to interesting physical 
implications.

Nevertheless, in the case of Barrow entropy one has an additional 
necessity for the varying behavior. Since  it arises 
from quantum-gravitational phenomena on the horizon 
structure, parametrized by the single exponent $\Delta$ \cite{Barrow:2020tzx}, 
where $\Delta=0$ corresponds to standard entropy and  $\Delta=1$ to maximal 
deviation,  it is   hard to justify why in the recent Universe one has   
significant quantum-gravitational phenomena in its horizon that could lead to 
significant deviation from standard, i.e. classical entropy. Hence, it is more 
natural to  consider that at early 
times these quantum 
phenomena are more intense and thus $\Delta$ is closer to 1, while as time 
passes they smooth out and $\Delta$ tends to its standard value 0. 
We mention here that such a scenario is still Barrow entropy, since $\Delta$ 
 does arise from quantum gravitational phenomena, however one goes beyond the 
first approximation of  \cite{Barrow:2020tzx} in which  these phenomena are 
constant throughout the Universe evolution, and examines more realistic 
scenarios in which      quantum gravitational phenomena 
are stronger at early times and smooth out as the system becomes larger.  

In summary, 
a scenario of Barrow entropy with varying exponent has even 
more justification than the usual energy-scale dependence of coupling constants.

 Since Barrow entropy  has been shown to 
lead to a very interesting cosmological phenomenology, in the present work we 
are interested in studying Barrow holographic dark 
energy with varying exponent. In particular, we impose various parametrizations 
for the time-dependence, which is equivalent to redshift-dependence, and we 
examine their effect on the cosmological evolution. The plan of the work is the 
following. In Section \ref{model0} we review standard Barrow holographic dark 
energy, and then in Section \ref{model1} we construct the extended scenario 
where the Barrow exponent is varying. In Section \ref{evolution} we impose 
specific   ans\"{a}tze  for the Barrow running exponent and we investigate 
their cosmological behavior. Finally, in Section  \ref{Conclusions}
we summarize our results.

\section{Standard Barrow holographic dark energy}         
\label{model0}

In this section we briefly review the scenario of Barrow holographic dark 
energy following \cite{Saridakis:2020zol}. Barrow entropy is given by  
\cite{Barrow:2020tzx}
\begin{equation}
\label{Barrsent}
S_B=  \left (\frac{A}{A_0} \right )^{1+\Delta/2}, 
\end{equation}
where $A$ is the standard horizon area  and $A_0$ the Planck area, while  the 
exponent $\Delta$ quantifies the quantum-gravitational deformation. 
Incorporating  (\ref{Barrsent}) in the definition of the standard holographic 
dark energy, expressed as the inequality $\rho_{DE} L^4\leq S$, with $L$ the 
horizon length, and imposing that $S\propto A\propto L^2 $ \cite{Wang:2016och} 
will result in 
\begin{equation}
\label{FRWTHDE}
\rho_{DE}={C} L^{\Delta-2},
\end{equation}
with ${C}$ a parameter with dimensions  $[L]^{-2-\Delta}$.
In   case where $\Delta=0$,   expression (\ref{FRWTHDE}) recovers   standard 
holographic dark energy $\rho_{DE}=3c^2 M_p^2 L^{-2}$ (here $M_p$ 
is the Planck mass), where ${C}=3 
c^2 M_p^2$ and $c^2$ the model parameter.

We focus on a flat homogeneous and isotropic Friedmann-Robertson-Walker (FRW)  
geometry with metric
\begin{equation}
\label{FRWmetric}
ds^{2}=-dt^{2}+a^{2}(t)\delta_{ij}dx^{i}dx^{j}\,,
\end{equation}
where $a(t)$ is the scale factor. Furthermore,   we consider that the 
universe is filled with the matter perfect fluid, as well as with the Barrow 
holographic dark energy \cite{Saridakis:2020zol}
\begin{equation}
\label{FRWTHDE2}
\rho_{DE}={C} R_h^{\Delta-2},
\end{equation}
where the
horizon length $L$ in (\ref{FRWTHDE}) is substituted by the future event 
horizon $R_h$\cite{Li:2004rb}, given by 
\begin{equation}
\label{FRWfuturehor}
R_h\equiv a\int_t^\infty \frac{dt}{a}= a\int_a^\infty \frac{da}{Ha^2},
\end{equation}
 with $H\equiv \dot{a}/a$ the Hubble parameter.
 We then obtain the two Friedmann equations  \begin{eqnarray}
\label{Fr1bFRW}
3M_p^2 H^2& =& \ \rho_m + \rho_{DE}    \\
\label{Fr2bFRW}
-2 M_p^2\dot{H}& =& \rho_m +p_m+\rho_{DE}+p_{DE},
\end{eqnarray}
with $\rho_m$ and $p_m$ 
the energy density and pressure of matter   and $p_{DE}$ the pressure of Barrow 
holographic dark energy.

The matter sector is conserved, namely it satisfies the continuity equation 
$
\dot{\rho}_m+3H(\rho_m+p_m)=0$.
Considering matter to be dust, namely imposing $p_m=0$, leads to  
  $\rho_m=\rho_{m0}/a^3$, with $\rho_{m0}$ the present matter 
energy density, i.e. at $a_0=1$ (in the following the subscript ``0'' denotes 
the  value of the corresponding quantity at present).
Finally, we focus on physically interesting observables such as the density 
parameters
 \begin{eqnarray}
 && \Omega_m\equiv\frac{1}{3M_p^2H^2}\rho_m
 \label{OmmFRW}\\
 &&\Omega_{DE}\equiv\frac{1}{3M_p^2H^2}\rho_{DE},
  \label{ODE}
 \end{eqnarray}
as well as on the effective dark-energy equation-of-state
parameter 
\begin{equation}
\label{wde}
w_{DE} \equiv \frac{p_{DE}}{\rho_{DE}}.
\end{equation}

Combining the aforementioned density parameters with equations (\ref{FRWTHDE2}) 
and (\ref{FRWfuturehor}), leads to  \cite{Saridakis:2020zol}
\begin{equation}\label{integrrelation}
\int_x^\infty \frac{dx}{Ha}=\frac{1}{a}\left(\frac{{C}}{3M_p^2H^2\Omega_{DE}}
\right)^{\frac{1}{ 2-\Delta}},
\end{equation}
 with  $x\equiv \ln a$. Additionally, substituting  $\rho_m=\rho_{m0}/a^3$ into 
(\ref{OmmFRW}), results to 
$\Omega_m=\Omega_{m0} H_0^2/(a^3 H^2)$, from which, using   the Friedmann 
equation   $\Omega_m+\Omega_{DE}=1$, 
we obtain
 \begin{equation}\label{Hrel2FRW}
\frac{1}{Ha}=\frac{\sqrt{a(1-\Omega_{DE})}}{H_0\sqrt{\Omega_{m0}}}.
\end{equation}

Inserting (\ref{Hrel2FRW}) into equation (\ref{integrrelation}) we acquire the 
expression
  \begin{equation}\label{inteion2FRW}
\int_x^\infty \frac{dx}{H_0\sqrt{\Omega_{m0}}}  
\sqrt{a(1-\Omega_{DE})}    =\frac{1}{a}\left(\frac{{C}}{
3M_p^2H^2\Omega_{DE}}
\right)^{\frac{1}{2-\Delta}}.
\end{equation}
Taking the derivative of  
(\ref{inteion2FRW}) with respect to $x=\ln a$   results to
  \begin{eqnarray}\label{Odediffeq0}
&&
\!\!\!\!\!\!\!\!\!\!\!\!\!
\frac{\Omega_{DE}'}{\Omega_{DE}(1-\Omega_{DE})}=\Delta+1+
Q
(1-\Omega_{DE})^{\frac{\Delta}{2(\Delta-2) } }  
 (\Omega_{DE})^{\frac{1}{2-\Delta } } 
e^{\frac{3\Delta}{2(\Delta-2)}x},
\end{eqnarray}
 with
   \begin{equation}\label{Qdef}
Q\equiv (2-\Delta)\left(\frac{{C}}{3M_p^2}\right)^{\frac{1}{\Delta-2}} 
\left(H_0\sqrt{\Omega_{m0}}\right)^{\frac{\Delta}{2-\Delta}},
\end{equation}
 and  where primes denote derivatives with respect to $x$.
 This is the differential equation determining the evolution of Barrow 
holographic dark energy, whose solution provides $\Omega_{DE}$.
Finally, since from    (\ref{FRWTHDE2}) we acquire
$\dot{\rho}_{DE}=(\Delta-2){C} R_h^{ \Delta-3} \dot{R}_h$, with $\dot{R}_h$ 
calculated using (\ref{FRWfuturehor}) as 
$\dot{R}_h=H  R_h-1$, and since $\dot{\rho}_{DE}+3H(\rho_{DE}+p_{DE})=0$, using 
(\ref{Odediffeq0}) we finally  extract \cite{Saridakis:2020zol}
\begin{equation}\label{wDEFRW}
w_{DE}=-\frac{1\!+\!\Delta}{3}
-\frac{Q}{3}
(\Omega_{DE})^{\frac{1}{2-\Delta } } (1\!-\!\Omega_{DE})^{\frac{\Delta}{
2(\Delta-2) } }
e^{\frac{3\Delta}{2(2-\Delta)}x}.
\end{equation}
 All the above expressions, for  $\Delta=0$ recover  standard holographic dark 
energy with $w_{DE}|_{\Delta=0} = - \frac{1}{3} 
-\frac{2}{3}\sqrt{\frac{3M_p^2 \Omega_{DE}}{C}}$  
\cite{Li:2004rb, Wang:2016och}. The scenario of Barrow holographic dark energy 
has interesting cosmological implications, and has been studied in detail in 
the literature \cite{Saridakis:2020zol,Anagnostopoulos:2020ctz,Nojiri:2021jxf,
Srivastava:2020cyk,Mamon:2020spa,Adhikary:2021xym,Chakraborty:2021uzp,
Huang:2021zgj,Luciano:2022ffn,Luciano:2022hhy,Rani:2021hvh,Luciano:2023wtx,
Oliveros:2022biu,Luciano:2022viz,Paul:2022doh,Jawad:2022qab,Boulkaboul:2023yks,
Sheykhi:2022fus,Feng:2023cbl}.

\section{Barrow holographic dark energy with varying exponent}
\label{model1}

In this section  we will investigate the cosmological scenario of Barrow 
holographic dark energy with varying exponent $\Delta$, namely we consider that 
  $\Delta \equiv  \Delta(x)$.
In this case, the analysis of the previous section remains the same up to 
(\ref{inteion2FRW}). However, taking its derivative requires to consider also 
the derivative of $\Delta$, which will yield $\Delta'$ terms in the equations. 
In particular, differentiating (\ref{inteion2FRW}) with 
respect to $x\equiv \ln a$, leads to 
\begin{eqnarray} 
\label{Odediffeq}
&&
\!\!\!\!\!\!\!\!\!\!\!\!\!\!\!
\frac{\Omega_{DE}'}{\Omega_{DE}\left(1-\Omega_{DE}\right)}= 
\sqrt{\Omega_{DE}} \left(\frac{{C}}{3M_p^2}\right)^{-\frac{1}{2}}    \left 
[\frac{P(1-\Omega_{DE})}{\Omega_{DE}}\right ]^{\frac{\Delta}{2(\Delta 
-2)}}(2-\Delta)+\Delta +1\nonumber\\
&& \ \ \ \ \ \ \ \ \  \ \ \  \ \ \   \ \   +\log{\left 
[\frac{P(1-\Omega_{DE})}{\Omega_{DE}}\right 
]}^{\frac{\Delta^{'}}{\Delta -2}},
 \end{eqnarray}
with 
\begin{equation}\label{Qparameter}
P=P(x)\equiv \frac{Ce^{3x}}{3 M_p^2H_0^2\Omega_{m0}}.
\end{equation}
 Equation (\ref{Odediffeq}) is the one that determines the evolution of Barrow 
holographic dark energy for dust matter in a flat universe, where primes denote 
derivatives with respect to $x$.

Additionally, let us calculate the equation-of-state parameter $w_{DE}$ for 
this  general scenario. We differentiate   (\ref{FRWTHDE2}), which leads 
to $\dot{\rho}_{DE}={C} R_h^{ \Delta-2}\left[\log{R}_h\,\dot{\Delta}+ 
\frac{\dot{R_h}(\Delta-2)}{R_h}\right] $, with  
$\dot{R}_h=H  R_h-1$.
Taking into account the  dark energy conservation equation 
$\dot{\rho}_{DE}+3H\rho_{DE}(1+w_{DE})=0$ and the aforementioned relations, we 
result to 
\begin{eqnarray} 
&&
(\Delta-2){C} 
\left(\frac{\rho_{DE}}{{C}}\right)^{\frac{\Delta-3}{\Delta-2}}
 \left[H  
\left(\frac{\rho_{DE}}{{C}}\right)^{\frac{1}{\Delta-2}}-1\right]\nonumber\\
&& 
+\rho_{DE}\dot{\Delta } 
\log
\left(\frac{\rho_{DE}}{C}\right)^{\frac{1}{\Delta-2}} +3H\rho_{DE}
(1+w_{DE})=0.
\end{eqnarray}
Hence, inserting $H$ from (\ref{Hrel2FRW}), and using    (\ref{ODE}) we finally 
obtain
\begin{equation}
 \label{wdeeq}
w_{DE}=-\frac{\Delta\! +\!1}{3}+\frac{(\Delta \!-\!2)\sqrt{\Omega_{DE}}}{3 } 
\left(\!\frac{{C}}{3M_p^2}\!\right)^{\!-\frac{1}{2}}\!\left 
[\frac{P(1\!-\!\Omega_{DE})}{\Omega_{DE}}\right ]^{\frac{\Delta}{2(\Delta 
\!-\!2)}}-\frac{\Delta^{'}}{3}\log{\left 
[\frac{P(1\!-\!\Omega_{DE})}{\Omega_{DE}} \right ]^{\frac{1}{2\!-\!\Delta}}}.
\end{equation}
In the case where $\Delta=const.$ we recover equation (\ref{wDEFRW}) of the 
previous section.

\section{Cosmological evolution}
\label{evolution}

In the previous section we extracted the necessary cosmological equations, 
which can describe the evolution in  the scenario at hand. In order 
to investigate in detail  the cosmological behavior, in what follows it is more 
convenient to use the redshift   $z$ instead of $x$ through the relation 
$x\equiv\ln a=-\ln(1+z)$. Hence, the derivative of a function $f$ in terms of 
$x$ easily becomes derivative in terms of $z$ through $f'=-(1+z) \frac{df}{dz}$.

We  will consider  several  ans\"{a}tze  for the Barrow running exponent, which 
will be given by the general expression $\Delta (z)=\alpha +\beta f(z)$, where 
$\alpha$, $\beta$ are constants and $f(z)$ a function of the redshift parameter 
$z$. Nevertheless, we impose  that the exponent $\Delta$   acquires values away 
from the standard $0$-value at early times, while as time passes 
it tends closer to the standard value $\Delta=0$.

We mention here that    we   parameterize  the way that quantum 
gravitational phenomena smooth out, i.e   $\Delta$ as a function of redshift, in 
a purely phenomenological manner, through various  functions. Definitely each 
of these functions at the fundamental level should correspond to a particular 
way that quantum gravitational phenomena behave.

Since  the differential equation (\ref{Odediffeq}) cannot be solved 
analytically,   we will elaborate it numerically. For   initial conditions we
impose 
$\Omega_{DE}(x=-\ln(1+z)=0)\equiv\Omega_{DE0}\approx0.7$ and therefore 
$\Omega_m(x=-\ln(1+z)=0)\equiv\Omega_{m0}\approx0.3$ in agreement with 
observations \cite{Planck:2018vyg}. 

\subsection{Linear case}

As a first example, let us consider the linear ansatz  $\Delta (z)=\alpha 
+\beta 
z$. Inserting it into equation 
(\ref{Odediffeq}) we acquire
\begin{eqnarray}\label{odedifflinear}
&&
\!\!\!\!\!\!\!\!\!\!\!\!\!
\frac{-(1+z) }{\Omega_{DE} 
\left(1-\Omega_{DE}\right)}  \frac{d \Omega_{DE}}{dz} = 
\sqrt{\Omega_{DE}} \left(\frac{{C}}{3M_p^2}\right)^{-\frac{1}{2}} \left 
[\frac{P(1-\Omega_{DE})}{\Omega_{DE}}\right ]^{\frac{\alpha +\beta z}{2(\alpha 
+\beta z -2)}}(2-\alpha -\beta z)\nonumber\\
&& \ \ \ \ \ \ \ \ \ \ \ \ \ \    \,\ \ \ \ 
\ \ \ \ \ \  \ \ 
+\log{\left [\frac{P(1-\Omega_{DE})}{\Omega_{DE}}\right ]}^{\frac{-\beta 
(1+z)}{\alpha +\beta  z -2}}+\alpha +\beta z +1, \hspace{2.8 em}
\end{eqnarray} 
and consequently equation (\ref{wdeeq}) becomes
\begin{eqnarray}\label{wdelinear}
&&
\!\!\!\!\!\!\!\!\!\!\!\!\!
w_{DE}=-\frac{\alpha +\beta z  +1}{3}+\frac{(\alpha +\beta z 
-2)\sqrt{\Omega_{DE}}}{3}\left(\frac{{C}}{3M_p^2}\right)^{-\frac{1}{2}}\left 
[\frac{P(1-\Omega_{DE})}{\Omega_{DE}}\right ]^{\frac{\alpha +\beta z}{2(\alpha 
+\beta z -2)}}\nonumber\\
&& \ \ \  
+\frac{(1+z)\beta}{3}\log{\left [\frac{P(1-\Omega_{DE})}{\Omega_{DE}}  \right 
]^{\frac{1}{2-\alpha -\beta z}}}.
\end{eqnarray}

\begin{figure}[!]
\centering
\includegraphics[width=6.5cm]{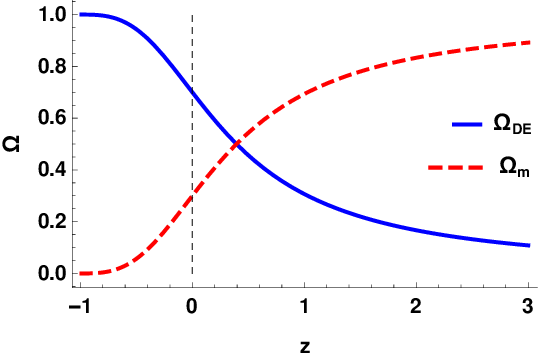}    \\                          
 \includegraphics[width=6.5cm]{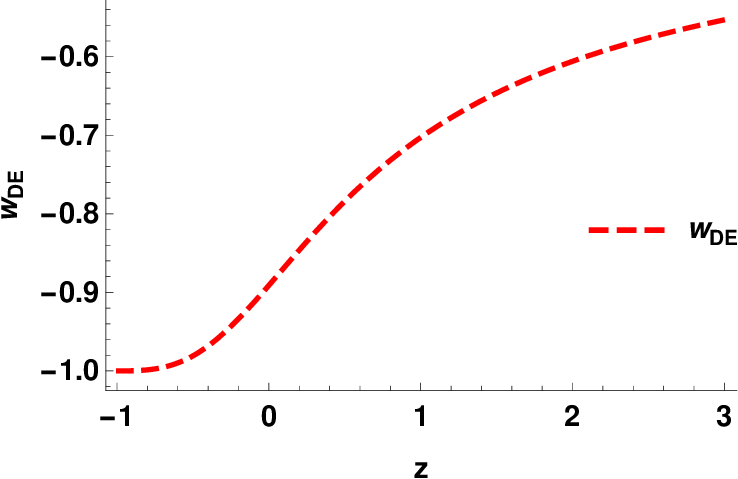} \\
\includegraphics[width=6.5cm]{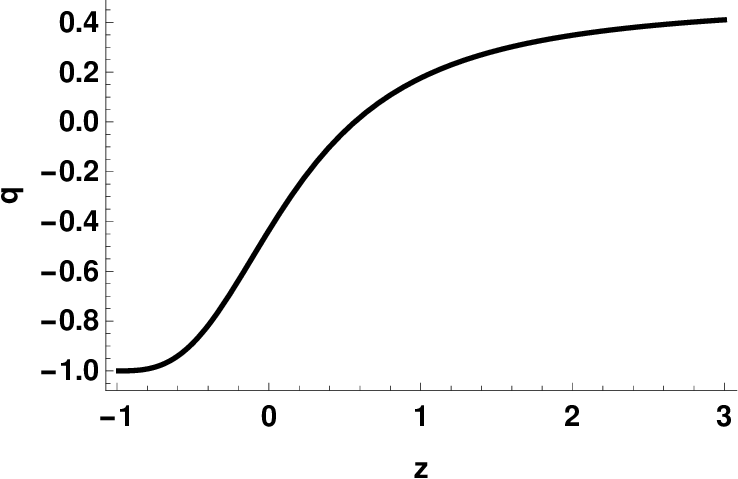}
\caption{\it{  Barrow holographic dark energy in the linear case where 
$\Delta 
(z)=\alpha +\beta z$.    {\bf{Upper graph}}:   The 
dark energy  density 
parameter  $\Omega_{DE}$ (blue-solid) and the matter density parameter 
$\Omega_{m}$ (red-dashed), as a function of the redshift $z$, for $C=1$ and
$\alpha=10^{-3}$, $\beta=10^{-4}$, in units where $M^{2}_{p} =1$.  {\bf{ 
Middle graph}}: The corresponding dark energy equation-of-state parameter 
$w_{DE}$. {\bf{Lower graph}}:  The corresponding deceleration parameter $q$. In 
all graphs we have set 
$\Omega_{DE}(x=-\ln(1+z)=0)\equiv\Omega_{DE0}\approx0.7$, 
in agreement with observations.
}}
\label{HDEOmegaslin}
\end{figure}

In the upper graph of Fig. \ref{HDEOmegaslin}, we depict the evolution of the 
dark energy and matter density parameters in terms of   redshift. As we can 
see, we acquire the usual thermal history of the universe, with the sequence of 
matter and dark energy epochs, and in the far future, namely at $z\rightarrow 
-1$, the universe results asymptotically to a complete dark energy dominated 
phase. Furthermore, in the middle graph we depict the evolution of the dark 
energy equation-of-state parameter, which is determined by     
(\ref{wdelinear}). As we observe, the value of $w_{DE}$ at present is around 
$-1$ in agreement with observations, lying in the quintessence regime, while 
in the future it asymptotically goes to the cosmological constant value. 
Lastly, in the lower graph we depict the deceleration parameter
  \begin{equation}
  \label{qdeccel}
q\equiv-1-\frac{\dot{H}}{H^2}=\frac{1}{2}+\frac{3}{2} w_{DE}
\Omega_{DE}.
\end{equation}
We can see the 
transition from deceleration to acceleration at  $z\approx 0.65$, 
in agreement with observational data. Note that, as required, the parameter 
$\beta$ should be suitably smaller than $\alpha$ in order for $\Delta$ to 
remain between 0 and 1. However, since there will be always a $z$ in which 
this will not be the case, it is clear that this ansatz is valid only at 
small redshifts. Hence, in the following we proceed to the investigation of 
 ans\"{a}tze with full applicability.

\subsection{CPL-like case}

Inspired by the Chevallier-Polarski-Linder (CPL) parameterization for the 
dark-energy equation of state \cite{Chevallier:2000qy,Linder:2002et}, we 
consider a  varying exponent of the form $\Delta (z)=\alpha +
\frac{\beta  z}{z+1}$, which can always be less than 1 for arbitrarily large 
redshifts.
  In this case  
substituting into   (\ref{Odediffeq}) we obtain
\begin{eqnarray}\label{odediffexp}
&&
\!\!\!\!\!\!\!\!\! 
\frac{-(1+z) }{\Omega_{DE}\left(1-\Omega_{DE}\right)} \frac{d \Omega_{DE}}{dz}= 
\sqrt{\Omega_{DE}} \left(\frac{{C}}{3M_p^2}\right)^{-\frac{1}{2}} \left 
[\frac{P(1-\Omega_{DE})}
{\Omega_{DE}}\right]^{\frac{\alpha (z+1) +
 \beta  z }{2(\alpha-2)(z+1) +
2\beta  z  }}\left[2\!-\!\alpha\! -\!
\frac{\beta  z}{z+1}\right]\nonumber\\
&& \ \ \ \ \ \ \ \ \ \ \ \ \ \  \ \ \ \ \ \ \   \,\ \ \ \ \ \ \ +\log{\left 
[\frac{P(1-\Omega_{DE})}{\Omega_{DE}}\right ]}^{-\frac{\beta }{(\alpha-2)  
(z+1) 
+
 \beta  z   }}+\alpha +
\frac{\beta  z}{z+1}+1,\hspace{2.8 em}
\end{eqnarray}
while equation (\ref{wdeeq})   gives
\begin{eqnarray}\label{wdeexp}
&&
\!\!\!\!\!\!\!\!\!\! 
w_{DE}=-\frac{\alpha +
\frac{\beta  z}{z+1} +1}{3}+\frac{\left(\alpha +
\frac{\beta  z}{z+1}
-2\right)\sqrt{\Omega_{DE}}}{3}\left(\frac{{C}}{3M_p^2}\right)^{-\frac{1}{2}}
\left 
[\frac{P(1-\Omega_{DE})}{\Omega_{DE}}\right ]^{\frac{\alpha (z+1) +
 \beta  z }{2(\alpha-2)(z+1) +
2\beta  z  }}\nonumber\\
&& \ \ \ \  \
-\frac{ \beta  }{3(z+1)}\log{\left 
[\frac{P(1-\Omega_{DE})}{\Omega_{DE}} \right ]^{\frac{z+1}{ (2-\alpha)(z+1) -
 \beta  z  }}} .
\end{eqnarray}
The behavior of dark energy and matter density parameters, of the deceleration 
parameter, and of the dark-energy equation-of-state parameter, is similar to 
the one of the linear case. Nevertheless, although very efficient in describing 
the past Universe evolution,  the CPL-like 
parametrization is not suitable for the future evolution, since the Barrow 
exponent will become larger than 1 or smaller than 0, and hence in the 
following we investigate more realistic cases.

\subsection{Exponential case}

One ansatz that can be suitable for all redshifts  is the  
exponential one, namely $\Delta (z)=\alpha +\beta e^{-\lambda z}$, where 
$\lambda$ is a constant. In this case  
substituting into   (\ref{Odediffeq}) we obtain
\begin{eqnarray}\label{odediffexp}
&&
\!\!\!\!\!\!\!\!\!\!\!\!\!
\frac{-(1+z) }{\Omega_{DE}\left(1-\Omega_{DE}\right)} \frac{d \Omega_{DE}}{dz}= 
\sqrt{\Omega_{DE}} \left(\frac{{C}}{3M_p^2}\right)^{-\frac{1}{2}} \left 
[\frac{P(1-\Omega_{DE})}
{\Omega_{DE}}\right]^{\frac{\alpha +\beta e^{-\lambda z}}{2(\alpha +\beta 
e^{-\lambda z} -2)}}[2-(\alpha +\beta e^{-\lambda z})]\nonumber\\
&& \ \ \ \ \ \ \ \ \ \ \ \ \ \  \ \ \ \ \ \ \   \,\ \ \ \ \ 
+\log{\left [\frac{P(1-\Omega_{DE})}{\Omega_{DE}}\right ]}^{\frac{(1+z)\beta 
\lambda e^{-\lambda z}}{\alpha +\beta e^{-\lambda z} -2}}+\alpha +\beta 
e^{-\lambda z}+1,\hspace{2.8 em}
\end{eqnarray}
while equation (\ref{wdeeq})   gives
\begin{eqnarray}\label{wdeexp}
&&
\!\!\!\!\!\!\!\!\!\! 
w_{DE}=-\frac{\alpha +\beta e^{-\lambda z} +1}{3}+\frac{(\alpha +\beta 
e^{-\lambda z} 
-2)\sqrt{\Omega_{DE}}}{3}\left(\frac{{C}}{3M_p^2}\right)^{-\frac{1}{2}}\left 
[\frac{P(1-\Omega_{DE})}{\Omega_{DE}}\right ]^{\frac{\alpha +\beta e^{-\lambda 
z}}{2(\alpha +\beta e^{-\lambda z} -2)}}\nonumber\\
&& \ \ \ \  \
-\frac{(1+z)\beta \lambda e^{-\lambda z}}{3}\log{\left 
[\frac{P(1-\Omega_{DE})}{\Omega_{DE}} \right ]^{\frac{1}{2-\alpha -\beta 
e^{-\lambda z}}}} .
\end{eqnarray}

\begin{figure}[!]
\centering
\includegraphics[width=6.5cm]{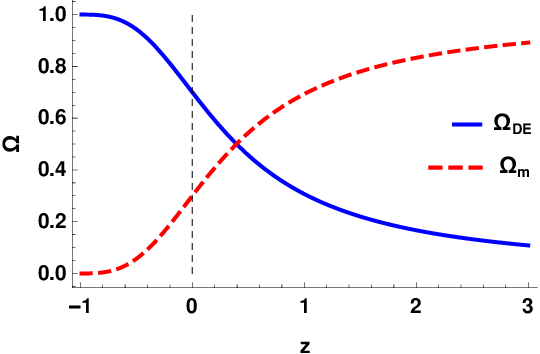}    \\                           
\includegraphics[width=6.5cm]{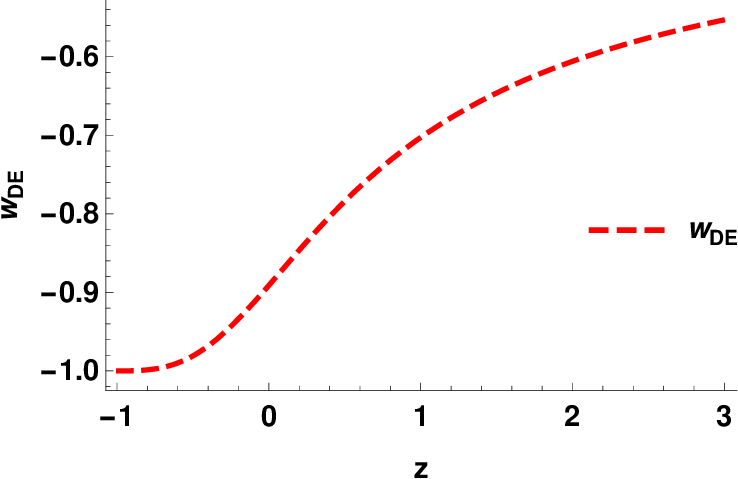} \\
\includegraphics[width=6.5cm]{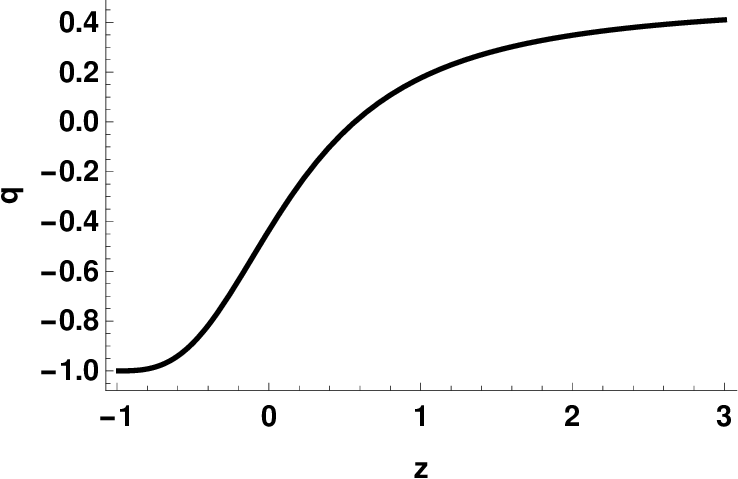}
\caption{\it{ Barrow holographic dark energy in the exponential case 
with $\Delta (z)=\alpha +\beta 
e^{-\lambda z}$.   {\bf{Upper graph}}:  The dark energy  density parameter 
$\Omega_{DE}$ (blue-solid) and the matter density parameter $\Omega_{m}$ 
(red-dashed), as a function of the redshift $z$, for $C=1$ and
$\alpha=10^{-3}$, $\beta=10^{-4}$, $\lambda =5\cdot 10^{-5}$, in units where 
$M^{2}_{p} =1$. {\bf{ Middle graph}}: The corresponding dark energy 
equation-of-state parameter $w_{DE}$. {\bf{Lower graph}}:  The corresponding 
deceleration parameter $q$. In all graphs we have set 
$\Omega_{DE}(x=-\ln(1+z)=0)\equiv\Omega_{DE0}\approx0.7$, in agreement with 
observations.
}}
\label{HDEOmegasexp}
\end{figure}

Fig. \ref{HDEOmegasexp}  shows the evolution of the dark energy and matter 
density parameters, and similarly to the linear case we acquire the usual 
thermal history of the universe. Additionally, from the middle graph we 
observe that $w_{DE}$ lies in the quintessence regime and its present value is 
around $-1$ according to observations. Finally, in the lower graph we 
depict the deceleration parameter $q$, where the transition from deceleration 
to acceleration in this case happens at $z\approx 0.65$.

\subsection{Hyperbolic Tangent case}

One ansatz that can be very suitable for the purpose of Barrow holographic dark 
energy with varying exponent is the hyperbolic tangent one, since in this case 
we can immediately bound $\Delta$ between 0 and 1 for all redshifts, obtaining 
easily a case  where $\Delta(z)$ is 1 at early times while it becomes 0 at 
intermediate and late times, as well as in the future.

\begin{figure}[ht]
\centering
\includegraphics[width=6.5cm]{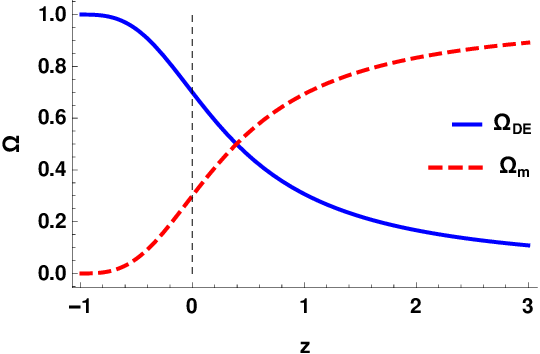}    \\                          
 \includegraphics[width=6.5cm]{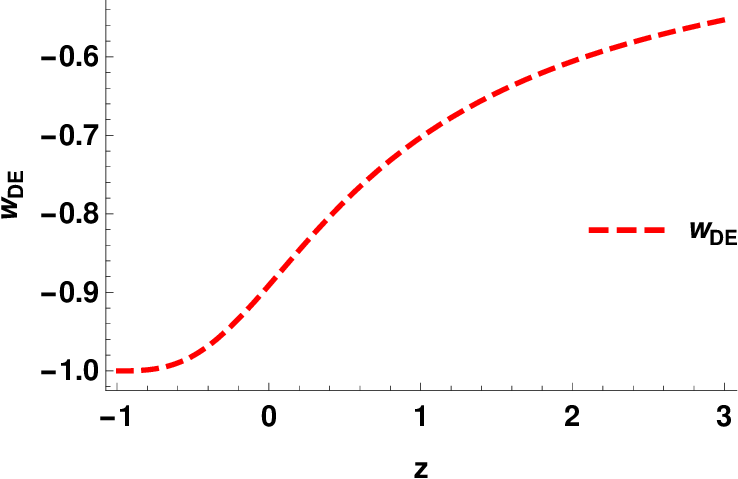} \\
\includegraphics[width=6.5cm]{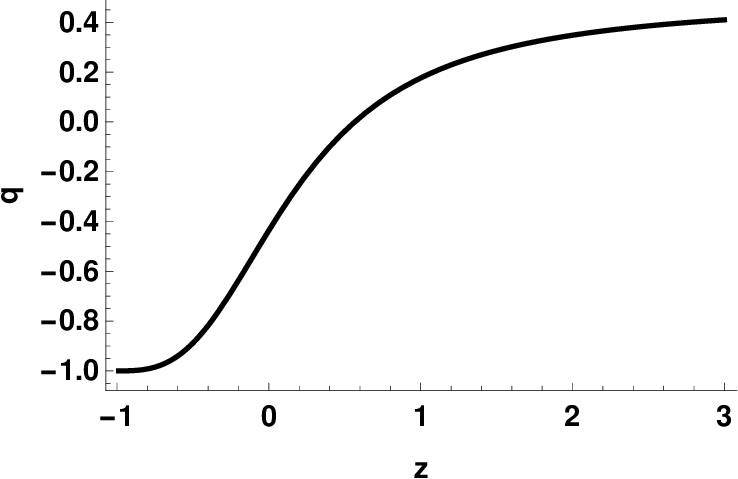}
\caption{\it{  Barrow holographic dark energy in the Hyperbolic Tangent 
case with $\Delta (z)=\alpha+ \beta \tanh({\gamma z})$, for  
$\alpha=\beta=\frac{1}{2}$. {\bf{Upper graph}}:  The 
dark energy  
density parameter $\Omega_{DE}$ (blue-solid) and the matter density parameter 
$\Omega_{m}$ (red-dashed), as a function of the redshift $z$, for $C=1$ and
$\gamma=0.001$, in units where $M^{2}_{p} =1$. {\bf{ Middle graph}}: The 
corresponding dark energy equation-of-state parameter $w_{DE}$. {\bf{Lower 
graph}}:  The corresponding deceleration parameter $q$. In all graphs we have 
set $\Omega_{DE}(x=-\ln(1+z)=0)\equiv\Omega_{DE0}\approx0.7$, in agreement with 
observations.
}}
\label{HDEOmegastanh}
\end{figure}

We consider  the form $\Delta (z)=\alpha+ \beta \tanh({\gamma z})$. Inserting 
  into equation (\ref{Odediffeq}) yields
\begin{eqnarray}\label{Odediffeqtanh}
&&
\!\!\!\!
\frac{-(1+z) }{\Omega_{DE}\left(1-\Omega_{DE}\right)}  \frac{d 
\Omega_{DE}}{dz} = 
\sqrt{\Omega_{DE}} \left(\frac{{C}}{3M_p^2}\right)^{-\frac{1}{2}} \left 
[\frac{P(1-\Omega_{DE})}{\Omega_{DE}}\right 
]^{\frac{ \alpha+ \beta \tanh({\gamma z})}{  2[\alpha+ \beta \tanh({\gamma z}) 
-2)]}}[2-\alpha- \beta \tanh({\gamma z})]
\nonumber\\
&& \ \ \ \ \ \ \ \ \ \ \ \ \  \ \ \  \  \ \ \ \ \ \ \ \ \ \ \ \ 
\ \
+\log{\left 
[\frac{P(1-\Omega_{DE})}{\Omega_{DE}}\right 
]}^{\frac{-(1+z)\beta\gamma \text{sech}^{2}({\gamma z})}{\alpha+ \beta 
\tanh({\gamma z})
-2}}
+\alpha+ \beta \tanh({\gamma z}) +1,\hspace{2.8 em}
\end{eqnarray}
and
\begin{eqnarray}\label{wdetanh}
&&
\!\!\!\!\!\!\!\!\!\!\!\!\!\!
w_{DE}=\frac{[\alpha\!+\! 
\beta \tanh({\gamma z}) \!
-\!2]\sqrt{\Omega_{DE}}}{3}\left(\frac{{C}}{3M_p^2}\right)^{-\frac{1}{2}}
\!\left [\frac{P(1-\Omega_{DE})}{\Omega_{DE}}\right 
]^{\frac{ \alpha+ \beta \tanh({\gamma z})}{  2[\alpha\!+\! \beta \tanh({\gamma 
z}) 
-2)]}}\nonumber\\
&& \ \ \  \,
-\frac{\alpha\!+\! \beta \tanh({\gamma z})\! +\!1}{3} 
+\frac{(1+z)\beta\gamma \,\text{sech}^{2}({\gamma z})}{3}\log{\left 
[\frac{P(1-\Omega_{DE})}{\Omega_{DE}} \right ]^{\frac{1}{2\!-\!\alpha\! -\! 
\beta\tanh{\gamma z}}}}.
\end{eqnarray}
In the following we set $\alpha=\beta=\frac{1}{2}$ which fixes the largest 
value of $\Delta(z)$ to 1 and the smallest to 0, in consistency with Barrow 
entropy. Hence the only model parameter is $\gamma$, which determines how fast 
the transition from 1 to 0 takes place.

\begin{figure}[h!]
\centering
\includegraphics[width=10cm]{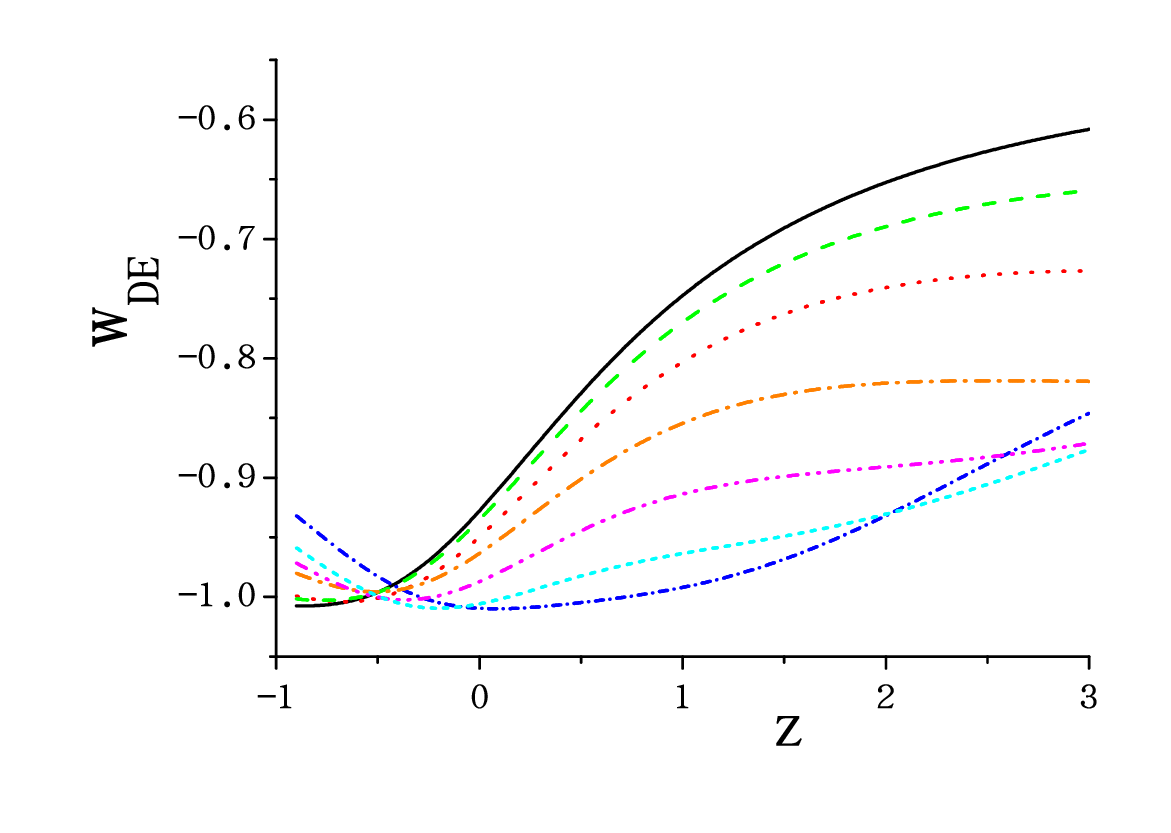}                           
\caption{\it{ The evolution of the dark-energy equation-of-state parameter for 
Barrow holographic dark energy in the Hyperbolic Tangent 
case with $\Delta (z)=\alpha+ \beta \tanh({\gamma z})$, for  
$\alpha=\beta=\frac{1}{2}$. The various curves correspond to $\gamma=0.01$
(black solid),  $\gamma=0.05$
(green dashed),  $\gamma=0.1$
(red dotted), $\gamma=0.2$
(orange dashed-dotted), $\gamma=0.3$
(magenta dashed-dotted-dotted), $\gamma=0.4$
(cyan short-dashed), and $\gamma=0.5$
(blue short dashed-dotted), in units where $M^{2}_{p} =1$. In all cases we have 
imposed $\Omega_{DE}(x=-\ln(1+z)=0)\equiv\Omega_{DE0}\approx0.7$ at present, in 
agreement with observations.
}}+
\label{wdeeffect}
\end{figure}

In the upper graph of Fig. \ref{HDEOmegastanh}  we depict the behavior of the 
dark energy and matter density parameters. As we can see, we acquire the  
sequence of matter and dark energy epochs, and in the far future  the universe 
results asymptotically to a complete dark-energy dominated phase. Furthermore, 
the dark energy equation-of-state parameter, in this specific example lies 
  in the quintessence regime, while at present times and in the future it 
is around $-1$.  Lastly,  the deceleration parameter $q$ experiences    the 
transition from deceleration to acceleration   at around $z\approx 0.65$.

Since the hyperbolic tangent form is the most realistic case, we explore it 
further. In particular, we are interested in examining what is the effect of 
the single parameter $\gamma$ on the dark-energy equation-of-state parameter.
In  Fig. \ref{wdeeffect}, we depict the evolution of $w_{DE}$, for various 
values of the $\gamma$. A general observation is that for increasing 
$\gamma$  the value of $w_{DE}$ decreases. Additionally, one can see 
that for some cases it can experience the phantom-divide crossing. 
Interestingly 
enough, in the future $w_{DE}$ can either tend to $-1$ or even start increasing 
again. These features reveals the capabilities of the model, and the advantages 
of Barrow holographic dark energy with running exponent comparing to standard 
Barrow holographic dark energy. 

We close this subsection by  confronting the model of the hyperbolic 
tangent case at hand with Supernovae 
type Ia (SNIa) and Cosmic Chronometer data.
For SNIa, it is well established that the relation between the apparent and 
absolute magnitudes follows  
 \begin{equation}
2.5 \log\left[\frac{L}{l(z)}\right] = \mu \equiv m(z) - M = 5 
\log\left[\frac{d_L(z)_{\text{obs}}}{Mpc}\right]  + 25~,
 \end{equation}
 with $l(z)$ and  $m(z)$ the apparent luminosity and apparent 
magnitude, and  $L$ and $M$    the absolute luminosity and 
magnitude, respectively, while $d_L(z)_{\text{obs}}$ is the luminosity distance.
Moreover, the theoretical expression for the luminosity distance is given by
\begin{equation}
d_{L}\left(z\right)_\text{th}\equiv\left(1+z\right)
\int^{z}_{0}\frac{dz'}{H\left(z'\right)}~.
\end{equation}
 \begin{figure}[h]
	\centering
\includegraphics[width=0.58\textwidth]{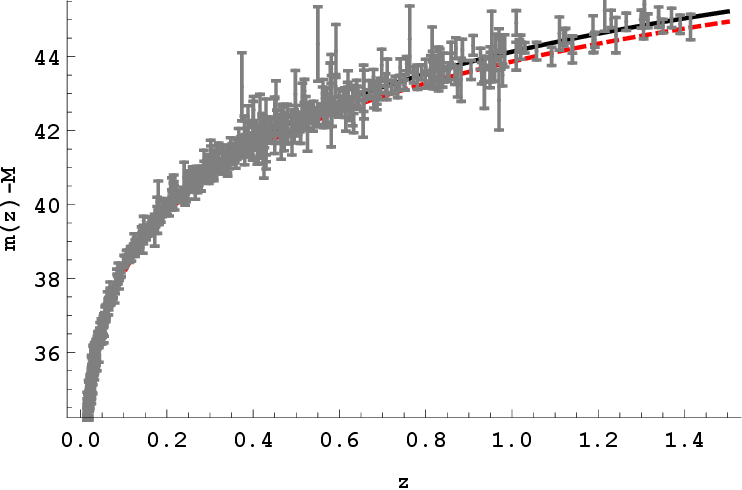}
	\caption{{\it{The apparent minus absolute magnitude 
	 predicted theoretically in the Hyperbolic Tangent case (red - dashed) with 
$\alpha=\beta=1/2$    and with 
$\gamma=0.001$, in units where $M_p^2=1$,  on top of 
the Pantheon SNIa data points from
\cite{Pan-STARRS1:2017jku}.
For comparison  we depict the $\Lambda$CDM  curve  (black - solid) too.}}}
	\label{fdatafig}
\end{figure}

 \begin{figure}[h]
	\centering
\includegraphics[width=0.61\textwidth]{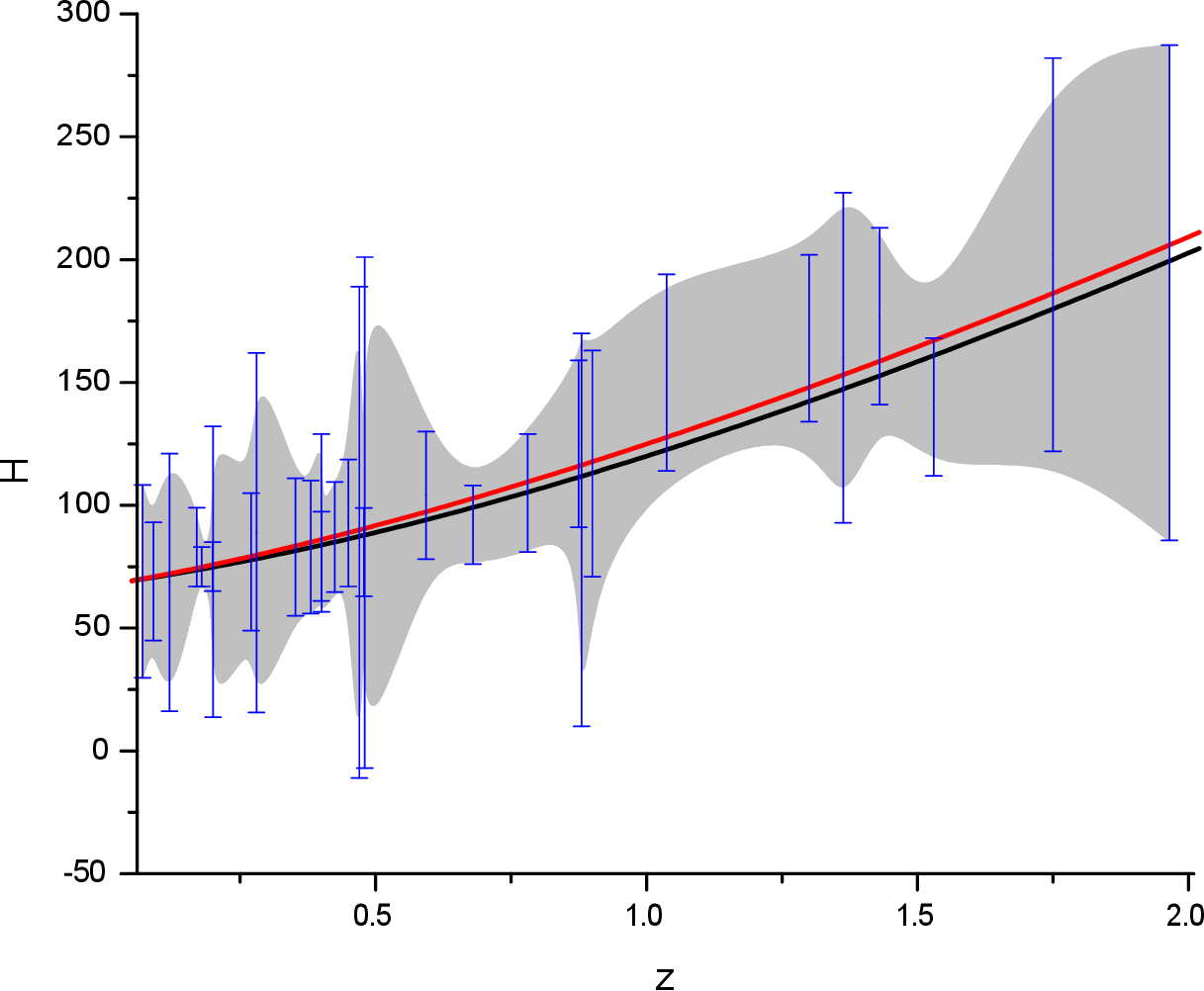}
	\caption{{\it{ The Hubble parameter $H(z)$ in units of Km/s/Mpc as 
a function of the redshift, for the hyperbolic tangent case  with $\alpha=\beta=1/2$   and 
with $\gamma=0.001$ (red-solid), in  $M_p^2=1$  units, on  top of the Cosmic Chronometers data 
points from \cite{Yu:2017iju} at $2\sigma$ confidence level. For comparison  we 
depict the $\Lambda$CDM  curve  (black - solid) too. We have imposed 
$\Omega_{m0}\approx0.31$.}} }
	\label{fdata2fig}
\end{figure}

Since the hyperbolic tangent case, as well as the $\Lambda$CDM 
scenario, provide predictions for $H(z)$, we compare the theoretically computed 
apparent minus absolute magnitudes with the binned Pantheon SNIa data from 
\cite{Pan-STARRS1:2017jku} in Fig. \ref{fdatafig}. As illustrated, the agreement 
is very good, with our model accurately reproducing the observed cosmological 
behavior.

In addition, the Cosmic Chronometer (CC) dataset, based on the 
measurement of $H(z)$ from the relative ages of passively evolving galaxies, 
allows us to compare the observed $H(z)$ values with our model predictions. In 
Fig. \ref{fdata2fig}, we show the comparison between the theoretical evolution 
of $H(z)$ in our model with a hyperbolic tangent form and the $\Lambda$CDM model 
with the CC data points from \cite{Yu:2017iju}, presented at a $2\sigma$ 
confidence level. The agreement is again very good, with our model reproducing 
the observed accelerating expansion for the parameter set $\{\alpha, \beta, 
\gamma\}= \{0.5, 0.5, 0.001\}$.

 In summary, there exist regions in the parameter space of our model 
that are able to reproduce the observed evolution of the Hubble function. This 
suggests that this model is a viable candidate.  
A more thorough evaluation, including likelihood analysis and model selection 
using complete cosmological datasets, will be presented in future work.

\subsection{Trigonometric functions case}

Finally, inspired by the oscillatory parametrizations of the dark-energy 
equation-of-state parameters \cite{Pan:2017zoh}, one could think of richer 
cases, where  the Barrow exponent oscillates around a mean positive value close 
to 0.
One such case could be the   $\Delta (z)=\alpha +\beta \sin{z}$ with 
$0<\beta<\alpha$, which leads to 
 \begin{eqnarray} \label{odediffsine}
&&
\!\!\!\!\!\!\!\!\!\!\!\!\!
\frac{-(1+z)\Omega_{DE}'}{\Omega_{DE}\left(1-\Omega_{DE}\right)}= 
\sqrt{\Omega_{DE}} \left(\frac{{C}}{3M_p^2}\right)^{-\frac{1}{2}} \left 
[\frac{P(1-\Omega_{DE})}{\Omega_{DE}}\right ]^{\frac{\alpha +\beta 
\sin{z}}{2(\alpha +\beta \sin{z} -2)}}(2-\alpha -\beta \sin{z})\nonumber\\
&& \ \ \ \ \ \ \   \ \ \ \ \  \ \ \ \ \ \,\
+\log{\left [\frac{P(1-\Omega_{DE})}{\Omega_{DE}}\right ]}^{\frac{-(1+z)\beta 
\cos{z}}{\alpha +\beta \sin{z} -2}}+\alpha +\beta \sin{z} +1, \hspace{3.2 em}
\end{eqnarray}
\begin{eqnarray}\label{wdesine}
&&
\!\!\!\!\!\!\!\! 
w_{DE}=-\frac{\alpha +\beta \sin{z} +1}{3}+\frac{(\alpha +\beta \sin{z} 
-2)\sqrt{\Omega_{DE}}}{3}\left(\frac{{C}}{3M_p^2}\right)^{-\frac{1}{2}}\left 
[\frac{P(1-\Omega_{DE})}{\Omega_{DE}}\right ]^{\frac{\alpha +\beta 
\sin{z}}{2(\alpha +\beta \sin{z} -2)}}\nonumber\\
&& \ \ \ \ \ \
+\frac{(1+z)\beta \cos{z}}{3}\log{\left [\frac{P(1-\Omega_{DE})}{\Omega_{DE}} 
\right ]^{\frac{1}{2-\alpha -\beta \sin{z}}}}.
\end{eqnarray}
 Alternatively, one could consider   $\Delta 
(z)=\alpha +\beta \cos{z}$,  obtaining 
\begin{eqnarray}\label{odediffcosine}
&&
\!\!\!\!\!\!\!\!\!\!\!\!\!
\frac{-(1+z)\Omega_{DE}'}{\Omega_{DE}\left(1-\Omega_{DE}\right)}= 
\sqrt{\Omega_{DE}} \left(\frac{{C}}{3M_p^2}\right)^{-\frac{1}{2}} \left 
[\frac{P(1-\Omega_{DE})}{\Omega_{DE}}\right ]^{\frac{\alpha +\beta 
\cos{z}}{2(\alpha +\beta \cos{z} -2)}}(2-\alpha -\beta \cos{z})\nonumber\\
&& \ \ \ \ \ \ \ \ \  \ \ \ \ \ \ \ \ 
\ \ 
+\log{\left [\frac{P(1-\Omega_{DE})}{\Omega_{DE}}\right ]}^{\frac{(1+z)\beta 
\sin{z}}{\alpha +\beta \cos{z} -2}}+\alpha +\beta \cos{z} +1, \hspace{2.8 em}
\end{eqnarray}
\begin{eqnarray}\label{wdecosine}
&&
\!\!\!\!\!\!\! 
w_{DE}=-\frac{\alpha +\beta \cos{z} +1}{3}+\frac{(\alpha +\beta \cos{z} 
-2)\sqrt{\Omega_{DE}}}{3}\left(\frac{{C}}{3M_p^2}\right)^{-\frac{1}{2}}\left 
[\frac{P(1-\Omega_{DE})}{\Omega_{DE}}\right ]^{\frac{\alpha +\beta 
\cos{z}}{2(\alpha +\beta \cos{z} -2)}}\nonumber\\
&& \ \ \ \ \ \  \,
-\frac{(1+z)\beta \sin{z}}{3}\log{\left [\frac{P(1-\Omega_{DE})}{\Omega_{DE}} 
\right ]^{\frac{1}{2-\alpha -\beta \cos{z}}}}
\end{eqnarray}
The behavior of dark energy and matter density parameters, of the deceleration 
parameter, and of the dark-energy equation-of-state parameter, is similar to 
the one of the previous models.

\section{Conclusions}
\label{Conclusions}

In this work we constructed Barrow holographic dark energy with varying 
exponent. This scenario is an extension of holographic dark energy with Barrow 
entropy, where the involved Barrow exponent, which quantifies the deviation 
from standard Bekenstein-Hawking entropy, is allowed to present a running 
behavior. Such an energy-scale-dependent behavior  is typical in  quantum field 
theory and quantum gravity under renormalization group considerations, however 
in the present scenario it has an additional justification, if not necessity, 
since in realistic cases one expects that  Barrow entropy quantum-gravitational 
effects to be stronger at early times and to smooth out and disappear at late 
times. Hence, in the cosmological framework of an expanding Universe, this 
would effectively generate Barrow-exponent with time-dependence.
 
After constructing the extended scenario we imposed specific   ans\"{a}tze  for 
the Barrow running exponent and we investigated their cosmological behavior. We 
started with the linear case, where we showed that we can recover the standard 
thermal history of the universe, with the sequence of 
matter and dark energy epochs. Additionally, we saw that the dark-energy 
equation-of-state parameter $w_{DE}$ lies in the quintessence regime and tends 
to $-1$ in the future, where the universe  results asymptotically to a complete 
dark energy dominated phase. Finally, from the behavior of the deceleration 
parameter we saw that the transition from deceleration to acceleration 
happens at  $z\approx 0.65$, in agreement with observations.

The linear ansatz is suitable only for small redshifts, since there will be 
always a redshift in which the Barrow exponent could exceed the theoretically 
determined  bounds. The CPL-like ansatz is applicable at all redshifts in the 
past, however is still not suitable for the future evolution. Hence, we 
proceeded to the examination of a more realistic case, namely the exponential 
ansatz. In this model, we obtained a similar behavior, with a 
sequence of matter and dark-energy epochs and the transition to acceleration at 
 $z\approx 0.65$ too.
 
 Nevertheless, the most suitable ansatz  for the purpose of Barrow holographic 
dark energy with varying exponent is the hyperbolic tangent one, since in this 
case we can easily bound Barrow exponent between 0 and 1 for all 
redshifts, obtaining a scenario  where $\Delta(z)$ is 1 at early times while it 
becomes 0 at intermediate and late times, as well as in the future. This 
scenario can also recover the thermal history of the Universe, and the recent 
transition to acceleration. Furthermore, we investigated the effect of the 
single parameter $\gamma$ on the   dark-energy 
equation-of-state parameter. We showed that for increasing 
$\gamma$  the value of $w_{DE}$ decreases, while  for some cases it can 
experience the phantom-divide crossing, and  that in the future $w_{DE}$ can 
either tend to $-1$ or  start increasing again. For completeness, we 
performed a basic  confrontation of the hyperbolic tangent model against the 
Supernovae type Ia (SnIa) 
and Cosmic Chronometers (CC) datasets, as a first
evidence that they are viable and consistent with observations. Lastly, we also
presented  the case where the running Barrow exponent has a
trigonometric form.

All these features reveal that  Barrow holographic dark energy with varying 
exponent is not only theoretically more justified than the standard, 
constant-exponent case, but it leads to richer cosmological behavior too.
 It would be interesting to perform a full observational confrontation with 
data from Supernovae type Ia (SNIa), Baryonic Acoustic Oscillations (BAO),   
Cosmic 
Microwave Background (CMB), and Cosmic Chronometers (CC) probes, in order to  
construct the likelihood contours and
extract constraints on the involved parameters, and apply information criteria 
in order to compare the scenario with the concordance $\Lambda$CDM cosmology. 
Such a full observational analysis is left for a future project.

\acknowledgments
 M.P. is supported by the Basic Research program of the National  Technical 
University of Athens (NTUA, PEVE) 65232600-ACT-MTG: Alleviating Cosmological 
Tensions Through
Modified Theories of Gravity. The authors acknowledge the contribution of the 
LISA
CosWG and of the COST Action CA21136 “Addressing observational tensions in 
cosmology with
systematics and fundamental physics (CosmoVerse)”.


\bibliographystyle{JHEP} 
\bibliography{references}

\end{document}